\documentclass[aps,preprint,preprintnumbers,amsmath,amssymb]{revtex4}
\def\be{\begin{equation}}
\def\ee{\end{equation}}
\def\beq{\begin{eqnarray}}
\def\eeq{\end{eqnarray}}
\def\n{\nonumber}

\begin{document} \openup8pt 
\preprint{smw-eqp-03-10}
\title{Explaining the equality of inertia and gravitational mass}
\author{Sanjay M Wagh}
\affiliation{Central India Research
Institute, \\ 34, Farmland, Ramdaspeth, Nagpur 440 010, India \\
E-mail: waghsm.ngp@gmail.com}

\begin{abstract}
The equality of the inertia and the gravitational mass of a body
is explained in a very general manner. We also motivate this
explanation by providing analogous examples.
\end{abstract}

\date{March 28, 2010}
\maketitle

Galileo's experiments suggested \cite{galileo} to him the uniform
rectilinear motion of a body to be its {\em natural\/} or {\em
inertial state of motion}. He had imagined the inertia of a body
as its opposition to a change in its natural state of motion, and
had conceptualized such a change to be due only to its interaction
with another body.

Newton completed \cite{principia} Galileo's conceptual framework
with his famous Three Laws of Motion. With an action-at-a-distance
interaction of bodies, Newton explained Kepler's laws of planetary
motion by assuming gravitational interaction of a planet and the
Sun. Then, the gravitational mass of a body is its source property
appearing in the postulated force of gravity, which it can be
assumed to exert on {\em all\/} other bodies.

Let us consider a body $A$ of gravitational mass $m$ and inertia
$\mathcal{I}$. Following Newton, we assume a force $\vec{F}$, of
the gravity of another body $B$ and acting on the body $A$, to be
proportional to the gravitational mass $m$ of the body $A$, {\em
ie}, $\vec{F}\propto m$ or $\vec{F} = -\,m\,\aleph\,\hat{r}$.
Here, $\aleph$ is a proportionality factor and $\hat{r}$ is a
radially outwardly directed unit vector along the line joining the
two bodies with the origin at the body $B$.

This force $\vec{F}$ causes the acceleration $\vec{a}$ of the body
$A$ towards the body $B$. Newton's second law of motion now
asserts that $\vec{F}=\mathcal{I}\,\vec{a}$. Then, we have $
\vec{F} = \mathcal{I}\,\vec{a} = -\,m\,\aleph \,\hat{r}$, and the
acceleration of the body $A$ is therefore given as \beq \vec{a}
=-\, \frac{m}{\mathcal{I}}\; \aleph\,\hat{r} \n \eeq Thus, the
magnitude of the acceleration of the body $A$ is proportional to
the ratio $m/\mathcal{I}$ of its gravitational mass and its
inertia. There is nothing within the Newtonian framework of
concepts (that is, of inertia, of action-at-a-distance force, of
gravitational mass, etc.) that compels this ratio to be unity,
{\em ie}, $m=\mathcal{I}$, for a body.

Consider now two such bodies, say, body 1 and body 2, for which
the proportionality factor $\aleph$ is the {\bf same}. Then, we
have \beq \vec{a}_1 =
-\,\frac{m_1}{\mathcal{I}_1}\,\aleph\,\hat{r} \hspace{.3in}
\mathrm{and}\hspace{.3in} \vec{a}_2 = -\,
\frac{m_2}{\mathcal{I}_2}\, \aleph\,\hat{r} \n \eeq Therefore, the
ratio of the magnitudes of these accelerations is \beq
\frac{a_1}{a_2} = \frac{m_1/\mathcal{I}_1}{m_2/\mathcal{I}_2} \n
\eeq Clearly, the accelerations of the two bodies would not be
same if the ratio $m/\mathcal{I}$ were not same for them, both.
Nothing within the Newtonian framework of concepts compels these
ratios to be equal for the two bodies under considerations.

Mach had then stressed \cite{mach} the lack of logically
compelling reasons for the inertia and the gravitational mass, two
conceptually different physical quantities, to be having {\em
exactly the same value\/} for a body, let alone for all the bodies
of Nature.

But, Galileo's experiments at the Leaning Tower of Pisa had shown
\cite{nagel} that the inertia and the gravitational mass of a body
are equal to a high degree of accuracy. Verifying this, and
explaining why this is so, have been certain issues, then.

In the present article, we show that this equality has a
remarkably simple and general explanation. To begin with, we
consider the following examples to illustrate fundamental
principles underlying this simple and very general explanation.

\begin{description} \item[{\bf Example One:}] To fix ideas, consider a body of ``volume'' $V$ and ``surface
area'' $A$. Let $V_{eq}=A\times\ell=V$ where $\ell$ is a certain
``hypothetical'' length. Then, for a sphere of radius $R$,
$\displaystyle{ \ell = R/3}$; for a cube of each edge of length
$a$, $\ell = a/6$; etc.

Now, $V_{eq}$ can have the meaning of the volume of ``another''
body, a {\em volume-equivalent body}, with a base of area $A$ and
a height $\ell$. By definition, $V$ and $V_{eq}$ are numerically
equal, then. But, we treat $V$, the ``volume of a body'', and
$V_{eq}$, the ``volume of a volume-equivalent body'', as being two
``different'' concepts.

When we evaluate another quantity, it makes no difference whether
we use the volume $V$ or the volume $V_{eq}$. For example, if
$\rho$ is the number density of particles in a body, then we would
obtain for the total number $N$ of particles in that body as
$N=\rho \times V$ and as $N'=\rho\times V_{eq}$. The values of $N$
and $N'$ obtained from these two expressions are bound to be
equal.

Furthermore, if $N=\rho V$ and $N=\rho' V_{eq}$, then no
``mystery'' is behind the equality of the values of $\rho$ and
$\rho'$ obtained from these expressions.

\item[{\bf Example Two:}] Consider now another example, that of the
specific heat of a substance in different systems of units.

Consider one kilogram of some substance of specific heat of, say,
$X$ k-cal/Kg/${}^o$C in the SI System of Units. Then, $X$
kilo-calories of heat are needed to change its temperature through
$1\,{}^o$C.

In the CGS System of Units, we have $1$ Kg of substance to be
equal to $1000$ gm of that substance, and $1$ k-cal is equal to
$1000$ cal. Then, the heat needed to change the temperature of $1$
gm of substance through $1$ ${}^o$C will be \[ X \times 1000
\times \frac{1}{1000} = X\;\mathrm{cal} \] The specific heat of
substance is $X$ cal/gm/${}^o$C in the CGS System of Units.

To any uninitiated mind, the sameness of the value of the specific
heat of any substance whatsoever in the aforementioned different
systems of units may appear surprising. It is however an artefact
of the following.

Notice that the unit of heat is defined in relation to the unit
mass of water and the unit change of temperature in all the
systems of units.

As has been defined, the specific heat of a substance is the ratio
of the ``heat needed to cause a unit change in the temperature of
the unit mass of the substance'' to the ``heat needed to cause a
unit change in the temperature of the unit mass of water'' in any
of these systems of units.

The specific heat, being this ratio, then has the same ``numerical
value'' in all the systems of units. This is not any
``surprising'' fact.
\end{description}

\begin{description} \item[{\bf Example Three:}] Now, for Coulomb's action-at-a-distance force, we have \beq
\vec{F} = \mathcal{I}\,\vec{a} \hspace{.3in}\mathrm{and}
\hspace{.3in} \vec{F} = q\,\left(-\,\Xi\,\hat{r}\right)
\hspace{.2in} \Longrightarrow \hspace{.2in} \vec{a}=
\frac{q}{\mathcal{I}} \left(-\,\Xi\, \hat{r}\right) \n \eeq with
the quantity $\left(-\,\Xi\,\hat{r}\right)$, $\Xi$ as a
proportionality factor, {\bf not} having the same dimensions as
those of $\vec{a}$. No equality of $q$ and $\mathcal{I}$ is then
compelled. \end{description}

We may now compare the situation of the examples discussed above
with that of the inertia and the gravitational mass of a body.
Consider then the relevant equations \beq \vec{F} =
\mathcal{I}\,\vec{a} \hspace{.3in}\mathrm{and} \hspace{.3in}
\vec{F} = m\,\left(-\,\aleph\,\hat{r}\right) \hspace{.2in}
\Longrightarrow \hspace{.2in} \vec{a}=\frac{m} {\mathcal{I}}
\left(-\,\aleph\, \hat{r}\right)\n \eeq Here, the dimensions of
$\left(-\,\aleph\,\hat{r}\right)$ and $\vec{a}$ are the {\bf
same}.

Consequently, the two quantities $\left(-\, \aleph\,
\hat{r}\right)$ and $\vec{a}$ ``refer'' to the same basic concept.
This is a case analogous to that of the volume $V$ and the volume
$V_{eq}$ in the first of the examples considered earlier. {\em The
values of $m$ and $\mathcal{I}$ will be compelled to be equal,
whenever the quantity $\left(-\, \aleph\, \hat{r}\right)$ is
``defined to have'' value equal to that of $\vec{a}$.}

Necessarily, the (operational part of the) definition of the
gravitational mass is based on, precisely, the equality of the
values of $\left(-\, \aleph\, \hat{r}\right)$ and of $\vec{a}$.
Then, why the inertia of a body is always equal to its
gravitational mass, even when the two are different physical
concepts, has the above simple explanation.

Now, the gravitational mass is an artefact of the assumption of
the action-at-a-distance force. However, no observation or
experiment imposes on us such a character for the force. The
concept of the gravitational mass of the body is therefore not an
``essential'' physical concept in the sense that its ``role'' is
similar to that of the volume $V_{eq}$ of a volume-equivalent body
in the first example considered above.

On the other hand, the concept of the inertia of a body is a
deduction, straightforwardly obtainable, from Galileo's
experiments. Therefore, the inertia needs to be considered as an
``essential'' concept in the sense that its ``role'' is similar to
that of the volume $V$ of a body in the first of the considered
examples.

But, Mach stressed \cite{mach} that, within Newton's theory, there
does not exist any principle to say that the inertia of a body
equals its gravitational mass, and this had been considered to be
one of the lacunae of Newton's theoretical framework.

In view of the explanation of this equality discussed in the
present work, no additional principle is required to explain the
equality of the inertia and the gravitational mass of a body.
Then, the lack of such a principle is ``not a shortcoming'' of the
Newtonian theory. But, contrary is what the equality of the
inertia and the gravitational mass of a body had us ``tricked''
into believing for long.

A striking ``internal asymmetry'' of Newton's theory based on the
postulated action-at-a-distance force but remains that the inertia
appears in the expression assumed for the force of gravity, and
not in the expressions for the other forces, like Coulomb's force.
This can be considered to be a certain shortcoming of Newton's
theory.

Conceivably, the Fundamental Theory of Physics should then be
based only on the ``contact'' interactions of bodies, and the
action-at-a-distance force could ``emerge' only as a suitable and
sometimes useful approximation within its framework. The
mathematical and the conceptual framework of the Universal Theory
of Relativity \cite{utr} then appears to provide for such a
theoretical possibility. \bigskip

\noindent {\bf Dedication} \bigskip

This work is dedicated to Professor P C Vaidya who passed away on
12th March 2010.


\end{document}